\pdfoutput=1
\documentclass[]{smule}
\usepackage[round,authoryear]{natbib}
\usepackage{hyperref}
\usepackage[colorinlistoftodos]{todonotes}
\usepackage{colortbl}
\usepackage{nicematrix}
\usepackage{amssymb}
\usepackage{amsmath}
\usepackage{booktabs}
\usepackage{adjustbox}
\usepackage{soul}
\usepackage[inline]{enumitem}
\usepackage{booktabs}
\usepackage{color}
\usepackage{xcolor}
\usepackage{bbding}
\usepackage{listings}
\usepackage{multicol}
\usepackage{xspace}
\usepackage{tablefootnote}
\usepackage{libertinus}

\makeatletter
\AtBeginDocument{

}
\makeatother

\title{Smule Renaissance Small: Efficient General-Purpose Vocal Restoration}

\author[1]{Yongyi Zang}
\author[1]{Chris Manchester}
\author[1]{David Young}
\author[1]{Ivan Ivanov}
\author[1]{Jeffrey Lufkin}
\author[1]{Martin Vladimirov}
\author[1]{PJ Solomon}
\author[1]{Svetoslav Kepchelev}
\author[3\dagger]{Fei Yueh Chen}
\author[2\dagger]{Dongting Cai}
\author[1]{Teodor Naydenov}
\author[1]{Randal Leistikow}

\affiliation[1]{Smule Labs}
\affiliation[2]{University of California, San Diego}
\affiliation[3]{University of Rochester}

\contribution[\dagger]{Work done during internship at Smule}

\abstract{
Vocal recordings on consumer devices commonly suffer from multiple concurrent degradations: noise, reverberation, band-limiting, and clipping. We present \emph{Smule Renaissance Small} (SRS), a compact single-stage model that performs end-to-end vocal restoration directly in the complex STFT domain. By incorporating phase-aware losses, SRS enables large analysis windows for improved frequency resolution while achieving 10.5$\times$ real-time inference on iPhone 12 CPU at 48 kHz. On the DNS 5 Challenge blind set, despite no speech training, SRS outperforms a strong GAN baseline and closely matches a computationally expensive flow-matching system. To enable evaluation under realistic multi-degradation scenarios, we introduce the \emph{Extreme Degradation Bench} (EDB): 87 singing and speech recordings captured under severe acoustic conditions. On EDB, SRS surpasses all open-source baselines on singing and matches commercial systems, while remaining competitive on speech despite no speech-specific training. We release both SRS and EDB under the MIT License.
}

\begin{document}

\maketitle

\section{Introduction}
Vocal recordings captured on consumer devices often exhibit background noise, reverberation, band-limiting, and clipping. These artifacts degrade perceived quality and hinder downstream digital signal processing (DSP). While a large body of work addresses individual distortions: e.g., denoising models~\citep{li2021two, wang2024mel}, dereverberation models~\citep{ernst2018speech, saito2023unsupervised}, vocal band-width extension models~\citep{iser2008bandwidth, wang2021towards, li2025audio}, or declipping models, real-world signals typically contain several distortions simultaneously, creating train–test mismatches that limit robustness and generalization.

Recent “clean-then-synthesize’’ systems attempt to address compound degradations by first predicting an intermediate representation and then vocoding~\citep{kong2020hifi}. VoiceFixer~\citep{liu2021voicefixer} predicts clean mel-spectrograms with a discriminative model and uses a GAN-trained vocoder for resynthesis; Resemble Enhance\footnote{\url{https://www.resemble.ai/introducing-resemble-enhance/}}
 similarly applies mel-domain enhancement followed by neural synthesis through latent flow matching. Although effective, two-stage designs increase computational cost, risk compounding stage-wise artifacts, and discard information when compressing to low-dimensional mel features.

We present \emph{Smule Renaissance Small} (SRS), a compact, single-stage, end-to-end vocal restoration model that operates directly in the complex short-time Fourier transform (STFT) domain. SRS uses a band-split generator to predict complex spectrograms, coupled with a temporal-convolutional backbone augmented by SwiGLU layers to model inter-band and real–imaginary channel dependencies. An auxiliary phase-optimization loss~\citep{li2025learning} enables training with large analysis windows, improving frequency resolution and temporal efficiency while maintaining phase consistency at synthesis time. To improve robustness to real-world mixtures of artifacts, we introduce a general-purpose corruption module that stochastically perturbs both magnitude and phase in addition to targeted degradations. This design avoids mel compression, reduces opportunities for error accumulation, and yields efficient CPU inference; on an iPhone~12 CPU, SRS runs at \mbox{10.5$\times$} real-time under 48 kHz.

Despite being trained without speech-specific data, SRS outperforms a stronger, GAN-trained baseline on the DNS~5 Challenge blind sets across standard metrics and is competitive with a latent flow-matching system that requires substantially more compute. A key impediment to comprehensive evaluation is the lack of realistic multi-degradation benchmarks. We therefore introduce \emph{Extreme Degradation Bench} (EDB), a curated set of singing and speech recordings collected under diverse, severe conditions. On EDB, SRS surpasses open-source baselines and matches commercial closed-source systems on singing while exceeding a GAN-trained open-source baseline on speech. We release both SRS~\footnote{\url{https://huggingface.co/smulelabs/Smule-Renaissance-Small}} and EDB~\footnote{\url{https://huggingface.co/datasets/smulelabs/ExtremeDegradationBench}} under the MIT License to facilitate reproducible research.

\section{Methods}
\label{sec:methods}

\subsection{Model Architecture}
\label{sec:model-arch}

\paragraph{Input representation.}
Given an input mixture waveform, we compute its complex-valued Short-Time Fourier Transform (STFT) to obtain $X\!\in\!\mathbb{R}^{B\times F\times T_s\times 2}$, where the final dimension separates real and imaginary components. The generator $G$ predicts a complex-valued estimate $\hat{X}=G(X)$ of identical shape, from which the enhanced waveform is recovered via inverse STFT: $\hat{y}=\mathcal{S}^{-1}(\hat{X})$.

The overall generator design is similar to~\cite{li2025apollo}:

\paragraph{Bandwise decomposition.}
We partition the frequency axis into $n_{\text{band}}$ contiguous sub-bands with mel-spaced boundaries. The integer width $bw_i$ of each band $i$ satisfies $\sum_{i=1}^{n_{\text{band}}} bw_i = F$.
For band $i$, we extract the corresponding frequency slice $X_i\in\mathbb{R}^{B\times bw_i\times T_s\times 2}$ and compute a per-frame power envelope:
\begin{equation}
p_i(t)=\sqrt{\textstyle\sum_{f\in \text{band } i}\big(\Re X_{f,t}\big)^2+\big(\Im X_{f,t}\big)^2+\varepsilon}\;\in\mathbb{R}^{B\times 1\times T_s},
\end{equation}
which serves both for normalization and as an explicit log-power feature input to the network.

\paragraph{Per-band feature extraction.}
Within each band, we normalize $X_i$ by dividing by $p_i(t)$ (broadcast across frequency bins and complex channels), then flatten the frequency and complex dimensions into $2\,bw_i$ channels. We concatenate this normalized representation with $\log p_i(t)$ to form $2\,bw_i+1$ input channels per time frame.
A lightweight stem consisting of RMSNorm followed by $1\!\times\!1$ pointwise convolution projects each band to a shared feature dimension $N$, yielding the initial hidden state:
\begin{equation}
H^{(0)}\in\mathbb{R}^{B\times n_{\text{band}}\times N\times T_s}.
\end{equation}

\paragraph{Band–Sequence block (repeated $L$ times).}
We design a modular processing block that couples cross-band spectral modeling with within-band temporal modeling:
\begin{itemize}
  \item \textbf{Cross-band attention.} We reshape $H^{(\ell-1)}$ to $(B\!\cdot\!T_s,\,N,\,n_{\text{band}})$ and apply multi-head self-attention along the band dimension. Queries and keys receive rotary position encoding (RoPE~\citep{su2024roformer}) based on band indices to preserve relative frequency ordering. We implement projections and output mixing via $1\!\times\!1$ pointwise convolutions, followed by a SwiGLU feedforward layer~\citep{shazeer2020glu}, with pre-norm RMSNorm and residual connections throughout.
  
  \item \textbf{Within-band temporal modeling.} We process each band independently using a depthwise-separable 1D ConvNeXT blocks~\citep{liu2022convnet} stack over the temporal dimension. The convolutions employ increasing dilation rates (e.g., $\{1,d,1\}$ where $d$ grows with network depth and is capped at a maximum value), followed by RMSNorm, a pointwise expansion with gated linear units (GLU), and a learned Layer scale parameter $\gamma$ for stable training.
\end{itemize}
The outputs of both pathways are summed with the block input via a long residual connection, maintaining the representation $H^{(\ell)}\!\in\!\mathbb{R}^{B\times n_{\text{band}}\times N\times T_s}$ across layers $\ell=1,\dots,L$.

\paragraph{Per-band synthesis and spectral reassembly.}
Each band employs a dedicated synthesis head comprising RMSNorm $\rightarrow$ $1\!\times\!1$ conv $\rightarrow$ SiLU $\rightarrow$ $1\!\times\!1$ conv $\rightarrow$ GLU, which maps the $N$-dimensional latent representation to $2\,bw_i$ output channels. These outputs are interpreted as real and imaginary components for each frequency bin within the band. We reshape each head's output to $(B,\,bw_i,\,T_s,\,2)$ and concatenate along the frequency axis to reconstruct the full complex spectrogram $\hat{X}\in\mathbb{R}^{B\times F\times T_s\times 2}$.

\paragraph{Computational complexity and design rationale.}
By restricting attention to the band axis rather than the frequency axis, the attention mechanism incurs cost $\mathcal{O}(B\,T_s\,n_{\text{band}}^2)$ instead of $\mathcal{O}(B\,T_s\,F^2)$. Temporal dependencies are captured through dilated depthwise convolutions with linear cost in $T_s$. This architectural decomposition enables effective spectral–temporal modeling with modest memory requirements, while RoPE preserves relative frequency relationships. The use of per-band synthesis heads allows for specialized reconstruction at different frequency ranges without requiring a computationally expensive global decoder.

\subsection{Training Setup}
\label{sec:training-setup}

\paragraph{Input–output notation.}
Let $x,y\in\mathbb{R}^{B\times T}$ denote the degraded mixture and clean target waveforms, respectively. We apply STFT analysis $\mathcal{S}$ and synthesis $\mathcal{S}^{-1}$ with window size 4096 and hop size 2048 to obtain spectral representations $X=\mathcal{S}(x)$ and $Y=\mathcal{S}(y)$.
The generator processes $X$ to predict $\hat{X}$, which is converted back to the time domain as $\hat{y}=\mathcal{S}^{-1}(\hat{X})$ for comparison with $y$. Note that this window and hop size configuration is substantially larger than those used in most existing systems; we find that incorporating phase-aware optimization losses enables acceptable synthesis quality even with this coarse temporal resolution.

\paragraph{Reconstruction objective.}
We employ a composite reconstruction loss combining time-domain, multi-resolution spectral magnitude, and phase-aware terms:
\begin{align}
\mathcal{L}_{\text{recon}}
&= \lambda_{\text{wav}} \,\|\hat{y}-y\|_1
 \;+\; \lambda_{\text{spec}} \,\big\|\,|\mathcal{S}(\hat{y})|-|\mathcal{S}(y)|\,\big\|_1
 \;+\; \lambda_{\text{omni}} \,\mathcal{L}_{\text{omni}}(\hat{X}, Y),
\end{align}
where $\mathcal{L}_{\text{omni}}$ denotes the phase-aware loss term introduced in~\cite{li2025learning}.

\paragraph{Adversarial objectives.}
We pair the generator $G$ with a multi-scale discriminator $D$ based on the Encodec architecture~\citep{defossez2022high}, comprising multi-period, multi-resolution STFT, and optional multi-band discriminator branches. We modify the original implementation by replacing weight normalization with spectral normalization for improved training stability.
Following the hinge loss formulation, the discriminator and adversarial losses are:
\begin{align}
\mathcal{L}_{D}
&= \textstyle\frac{1}{K}\sum_{k=1}^{K}
\Big( \mathbb{E}\,[\max(0,\,1 - D^k_{\phi_i}(y))]
      + \mathbb{E}\,[\max(0,\,1 + D^k_{\phi_i}(\hat{y}))] \Big),\\
\mathcal{L}_{\text{adv}}
&= -\textstyle\frac{1}{K}\sum_{k=1}^{K} \mathbb{E}\,[D^k_{\phi_i}(\hat{y})],
\end{align}
where $K$ denotes the number of discriminator branches. We additionally incorporate a normalized feature-matching loss:
\begin{align}
\mathcal{L}_{\text{fm}}
&= \textstyle\frac{1}{K}\sum_{k=1}^{K}\;\frac{1}{L_k}\sum_{\ell=1}^{L_k}
\frac{\big\|\phi^{k,\ell}(y)-\phi^{k,\ell}(\hat{y})\big\|_1}{\operatorname{mean}(|\phi^{k,\ell}(y)|)+\varepsilon},
\end{align}
where $\phi^{k,\ell}$ denotes the feature map at layer $\ell$ of discriminator $k$, and $L_k$ is the number of layers in that discriminator. The total generator loss combines all objectives:
\begin{equation}
\mathcal{L}_{G}=\mathcal{L}_{\text{recon}}+\lambda_{\text{adv}}\mathcal{L}_{\text{adv}}+\lambda_{\text{fm}}\mathcal{L}_{\text{fm}}.
\end{equation}

\paragraph{Optimization and regularization.}
We train both generator and discriminator using AdamW optimizers with shared weight decay and $\epsilon$ hyperparameters. The learning rate follows a linear warmup schedule followed by cosine decay. To ensure training stability, we apply global gradient norm clipping with a threshold of $1.0$ to both networks at each optimization step.

\subsection{Data}
\paragraph{Dataset collection.} 
Our training dataset consists of singing voice recordings collected in professional recording studios. Participants performed a standardized singing elicitation protocol~\footnote{Dataset manuscript is in preparation and will be released in a future publication.}, and recordings were subsequently processed by audio engineers to ensure quality. During each recording session, we captured audio from multiple microphones simultaneously; all microphone channels are utilized during training to increase data diversity.

\paragraph{Degradation simulation.} 
To train the model on diverse degradation scenarios, we apply a comprehensive augmentation pipeline. We simulate frequency-dependent degradation, reverberation using parametric room models, various clipping curves, and additive environmental and instrumental noise. Additionally, we introduce stochastic perturbations in both magnitude and phase domains by randomly masking frequency bins in the magnitude spectrogram and adding noise to the phase spectrogram. These phase perturbations are computed under randomly sampled STFT parameters drawn from the following sets: window sizes $\in\{512, 1024, 2048\}$ and hop sizes $\in\{256, 512, 1024\}$. Finally, we apply time-varying gain modulation by generating random noise, applying a lowpass filter to create a smooth gain envelope, and multiplying this envelope with the audio signal to simulate realistic volume fluctuations.

\section{Results}
\subsection{Objective Performance}
\begin{table}[h]
\centering
\begin{tabular}{cccccc}
\toprule
\multicolumn{1}{l}{}     & SIG           & BAK           & OVRL          & UTMOS         & Average       \\
\midrule
VoiceFixer               & 3.38          & 3.90          & 3.04          & 2.03          & 3.09          \\
Resemble Enhance         & \textbf{3.54} & \textbf{3.98} & \textbf{3.22} & \textbf{2.35} & \textbf{3.27} \\
SRS (Ours) & \underline{3.50} & \textbf{3.98} & \underline{3.18} & \underline{2.13} & \underline{3.20} \\
\bottomrule
\end{tabular}
\caption{Objective results on DNS 5 Challenge Blind Set. Best performances are in \textbf{bold} and second-best performances are \underline{underlined}.}
\label{tab:objective}
\end{table}

We evaluate SRS against two open-source systems on the DNS~5 Challenge~\citep{dubey2024icassp} blind set using predictor-based MOS metrics: DNSMOS~P.835~\citep{reddy2022dnsmos} \emph{SIG} (speech quality), \emph{BAK} (background intrusiveness), \emph{OVRL} (overall quality), and \emph{UTMOS}~\citep{saeki2022utmos}; higher is better. Table~\ref{tab:objective} reports per-metric scores and their unweighted mean.

SRS is top-2 on all metrics, tying for the best \emph{BAK} (3.98) and landing within 0.04 of the best system on both \emph{SIG} (3.50 vs.\ 3.54) and \emph{OVRL} (3.18 vs.\ 3.22). Relative to VoiceFixer, SRS improves by +0.12 \emph{SIG} (3.50 vs.\ 3.38), +0.08 \emph{BAK} (3.98 vs.\ 3.90), +0.14 \emph{OVRL} (3.18 vs.\ 3.04), and +0.10 \emph{UTMOS} (2.13 vs.\ 2.03), yielding a +0.11 gain in the average score (3.20 vs.\ 3.09). While Resemble Enhance attains the highest average (3.27), SRS is close behind at 3.20 ($\Delta=0.07$) despite being a single-stage model, supporting a favorable accuracy–efficiency trade-off.

\subsection{Extreme Degradation Bench (EDB)}
Current vocal restoration test sets primarily consist of blind subsets from various noisy speech datasets, which typically contain limited degradation types and severity levels. To address this limitation, we propose the Extreme Degradation Bench (EDB), a benchmarking dataset comprising 87 14-second mono 48 kHz audio clips captured under diverse degradation conditions. The dataset includes samples from the UCSB Cylinder Audio Archive~\footnote{\url{https://cylinders.library.ucsb.edu/}}, as well as audio recorded in challenging acoustic environments including public transport, airports, household settings, and outdoor sports venues. EDB contains both singing and speech content across multiple languages and regions, providing comprehensive coverage of real-world degradation scenarios.

To comprehensively benchmark current system performance, we compare our proposed SRS system against the aforementioned open-source baselines and two commercial closed-source systems: Adobe Enhance V2~\footnote{\url{https://podcast.adobe.com/en/enhance}, accessed October 11, 2025 via web interface} and Lark V2~\footnote{\url{https://ai-coustics.com/2025/07/29/lark-2-next-generation-reconstructive-speech-enhancement/}, accessed October 11, 2025 via official API endpoint}.

\subsubsection{Subjective Ranking}
We conducted a subjective ranking study in which participants evaluated 20 pairwise comparisons. For each comparison, participants heard the original degraded clip followed by outputs from two randomly selected systems, then indicated whether one system was superior or whether the systems were tied. The study ran for one week and collected responses from 34 participants. All submitted ratings were included in the analysis, even from participants who did not complete the full set of comparisons. We fitted a Bradley-Terry model to the pairwise comparison data to compute ELO scores for each system.

\begin{figure*}[h!]
  \centering
\includegraphics[width=\textwidth]{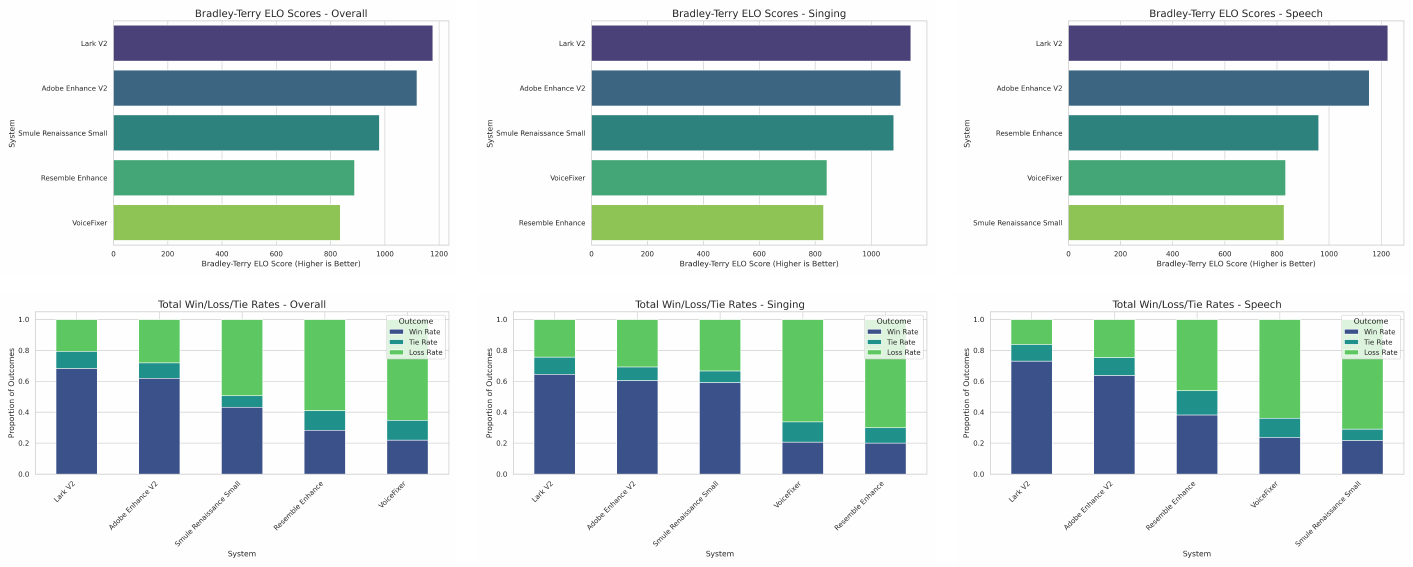}
\caption{Overall results for all systems on Extreme Degradation Bench.}
  \label{fig:overall-results}
\end{figure*}

Results are shown in Figure~\ref{fig:overall-results}. We observe that closed-source systems achieve strong performance across overall, singing, and speech categories, with Lark V2 consistently outperforming Adobe Enhance V2. Our proposed SRS system achieves the highest performance among open-source systems overall and attains performance comparable to closed-source systems on singing restoration. Notably, despite receiving no explicit training on speech data, SRS demonstrates performance on extremely degraded speech comparable to VoiceFixer, a larger GAN-based system specifically designed for speech enhancement.

\begin{figure*}[h!]
  \centering
\includegraphics[width=\textwidth]{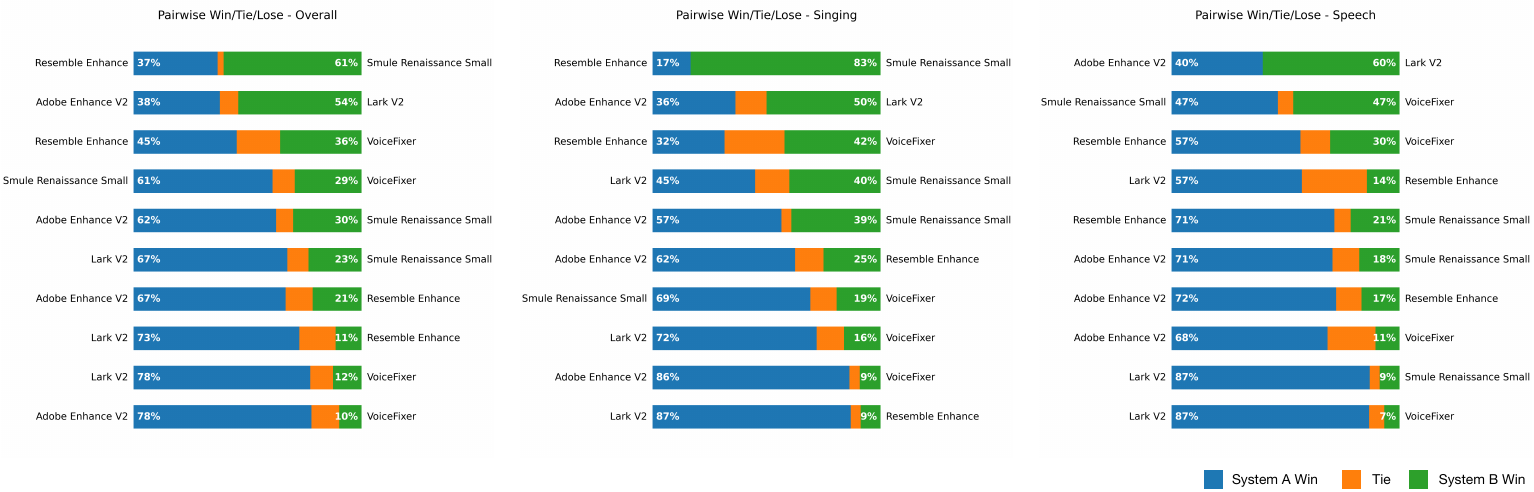}
\caption{Pairwise win/tie/lose results for all systems on Extreme Degradation Bench.}
  \label{fig:pairwise-results}
\end{figure*}

Figure~\ref{fig:pairwise-results} presents detailed pairwise win/tie/loss rates between all systems. The results align with expected performance hierarchies: closed-source systems generally outperform open-source alternatives, with Lark V2 exhibiting a modest but consistent advantage over Adobe Enhance V2. SRS outperforms all open-source systems in both overall and singing categories. Particularly noteworthy is that SRS shows only a small performance gap compared to Lark V2 on singing restoration tasks. While SRS demonstrates relatively weaker performance on speech restoration, it remains competitive with VoiceFixer—a GAN-based network trained specifically for speech—which corroborates the overall win/loss/tie rate observations.

\subsubsection{Bradley-Terry Model Validation} 
To verify the validity of our evaluation methodology, we assessed the goodness-of-fit of the Bradley-Terry model by computing R-squared, Mean Absolute Error (MAE), and Root Mean Squared Error (RMSE) between predicted and observed pairwise comparison outcomes.

\begin{table}[h!]
\centering
\begin{tabular}{lccc}
\toprule
Category & R² & MAE & RMSE \\
\midrule
Overall & 0.9540 & 0.0203 & 0.0243 \\
Speech & 0.9000 & 0.0298 & 0.0416 \\
Singing & 0.8171 & 0.0515 & 0.0591 \\
\bottomrule
\end{tabular}
\caption{Goodness-of-fit metrics for Bradley-Terry model across evaluation categories.}
\label{tab:bt-validation}
\end{table}

As shown in Table~\ref{tab:bt-validation}, all categories exhibit strong model fit, with R² values exceeding 0.8 and low prediction errors, confirming that the Bradley-Terry model provides a valid and reliable framework for ranking system performance in this evaluation.

\subsection{Mobile Device Performance}
To evaluate the practical deployment of our system, we benchmark inference performance on consumer iOS devices using a 10-second audio input at 48 kHz sampling rate. We test on two devices representing different generations: iPhone 12 (released 2020, mid-tier) and iPhone 14 Pro (released 2022, flagship), measuring latency on both CPU and GPU accelerators\footnote{Benchmarks conducted using Xcode's built-in profiling tools for MLPackage deployment.}.

\begin{table}[h]
\centering
\begin{tabular}{lccccc}
\toprule
Device & Compute & Median (s) & P90 (s) & Mean (s) \\
\midrule
iPhone 12 & GPU & 1.057 & 1.540 & 1.384 \\
iPhone 12 & CPU & 0.948 & 0.970 & 0.952 \\
iPhone 14 Pro & GPU & 0.631 & 0.795 & 0.780 \\
iPhone 14 Pro & CPU & 0.727 & 0.739 & 0.731 \\
\bottomrule
\end{tabular}
\caption{Inference latency on mobile devices for processing 10 seconds of audio. Median, 90th percentile (P90), and mean values computed over repeated runs.}
\label{tab:latency_mobile}
\end{table}

Results are presented in Table~\ref{tab:latency_mobile}. On the iPhone 12, CPU execution achieves faster and more consistent performance (0.948s median) compared to GPU (1.057s median). We attribute the slower GPU performance to cold start overhead between CPU and GPU on this older device architecture. Nevertheless, even on this five-year-old consumer device, the system achieves approximately 10.5$\times$ real-time processing speed~\footnote{Note that the system is not causal, and therefore cannot actually be used in real-time.}.

On the more recent iPhone 14 Pro, both accelerators demonstrate improved performance. The GPU achieves a median latency of 0.631s, corresponding to 15.8$\times$ real-time speed, while the CPU maintains competitive performance at 0.727s (13.7$\times$ real-time). The improved GPU performance on iPhone 14 Pro suggests that newer device architectures have reduced memory transfer bottlenecks, making GPU acceleration more effective.

These results demonstrate that our system achieves practical real-time performance across multiple generations of consumer mobile hardware, validating its deployment viability for on-device audio restoration applications.

\section{Conclusion}
\label{sec:conclusion}

We presented Smule Renaissance Small (SRS), a compact single-stage vocal restoration system operating directly in the complex STFT domain. By combining band-split architecture with phase-aware losses, SRS achieves effective restoration with 10.5$\times$ real-time inference on iPhone 12 CPU, demonstrating practical viability for on-device deployment. Our evaluations show that SRS outperforms strong GAN baselines and approaches more expensive flow-matching systems on DNS 5 Challenge—despite no speech training. On the Extreme Degradation Bench, SRS surpasses all open-source alternatives on singing and matches commercial systems, while remaining competitive on speech without speech-specific training.  We release both SRS and EDB under the MIT License to facilitate reproducible research in robust vocal restoration.

\clearpage
\newpage
\bibliographystyle{plainnat}
\bibliography{paper}

\end{document}